\documentstyle[preprint,aps,psfig,epsf,rotate]{revtex}
\begin{document}
\draft

\title{Ground-States of Two Directed Polymers}

\author{V. T. Pet\"aj\"a$^1$, M. J. Alava$^{1}$,
and H. Rieger$^{2}$}

\address{$^1$ Helsinki University of Technology, Laboratory of
Physics,\\ P.O.Box 1100, FIN-02015 HUT, Finland}
\address{$^2$ Theoretische Physik, Universit\"at des Saarlandes,
66041 Saarbr\"ucken, Germany}

\maketitle

\begin{abstract}
  Joint ground states of two directed polymers in a random medium are
  investigated. Using exact min-cost flow optimization the true two-line
  ground-state is compared with the single line ground state plus its first
  excited state. It is found that these two-line configurations are (for
  almost all disorder configurations) distinct implying that the true
  two-line ground-state is non-separable, even with 'worst-possible' initial
  conditions. The effective interaction energy between the
  two lines scales with the system size with the scaling exponents 
  0.39 and 0.21 in 2D and 3D, respectively.
\end{abstract}

\noindent {\it PACS \# \ \ 05.40.-a, 05.50+q, 05.70.Np, 75.50.Lk}
\newpage

\newcommand{\be}{\begin{equation}}
\newcommand{\ee}{\end{equation}}

\section{Introduction}

The physics of disordered systems has attracted a lot of attention due to the
discovery that the free energy of extended objects - lines, surfaces and so on
- has singular corrections because of the domination of zero-temperature or
ground-state effects \cite{halpin}. The paradigm of such systems is a directed
polymer in a random medium (DPRM). In this particular example the object
minimizes its energy which is determined by two competing forces: the elastic
energy cost of wandering on one hand and the energy gain using energetically
favorable pins in the environment on the other hand. The result is
super-diffusive behavior, and constrained energy fluctuations. The phase space
of the DPRM problem is very rich depending on the nature of the correlations
in the disorder and the dimensionality. In low enough dimensions the physics
is (at arbitrary temperatures) governed by the so-called zero-temperature
fixed point if the noise has weak enough correlations including the
uncorrelated case. The case with one transverse dimension becomes exactly
solvable in terms of the roughness and energy fluctuation exponents, due to a
mapping to the Kardar-Parisi-Zhang equation \cite{kardar}. The values are
$\zeta_2 = 2/3$ and $\theta_2 = 1/3$, which fulfill the exponent relation 
$2\zeta_d - 1 = \theta_d$. In the 3(=2+1)-dimensional case the roughness
exponent is approximately $0.62$.

In this paper we study the problem of two (not necessarily directed) polymers
in a joint random medium (TPRM) with mutual interactions
\cite{mezard,natter,tang,muk,kinzel,hwa} and focus on the repulsive strong
coupling limit, i.e.\ hard core interaction.  The work is related both to the
question of the physics of flux-lines in high-$T_c$-semiconductors in the low
field limit, and to the field-theoretical issues due to the importance of the
DP interaction energy.  The physics of the problem is in general very similar
to that of the one-line case \cite{oneline_num,oneline,mezard} but shows
interesting twists if one tries to understand the problem in the light of
individual, independent objects. In particular, we are going to consider by
numerical, {\it exact} min-cost flow optimization computations the difference
in energy between the TPRM problem, the single-line ground-state and the 'first
excited state'. The last one is given by adhering to a hierarchical picture,
in which the first polymer is first optimized given a disorder configuration,
and then the next one is added by applying a hard-core repulsion to the bonds
already taken up. The procedure gives us two energies to compare with the true
TPRM ground state energy $E_2$: the single line ground-state energy doubled,
$2E_1$, and the sum of the ground-state energy $E_1$ and the energy of the
first excited state $E_1'$ in the single line problem.  The two energy
differences, $E_2-2E_1$ and $E_2-E_1-E_1'$ define an interaction energy of the
two polymers. In an earlier paper Tang \cite{tang} studied the TPRM in
hierarchical lattices and in two dimensions with binary disorder. His main
conclusion was, for the physically more relevant real-space case, that the
probability for an interaction energy exactly equal to zero (with binary
disorder) decayed much faster than expected, the exponent being -2/3 instead
of the -1/3 expected based on single-DP geometric arguments. We study both the
interaction energies discussed above. We also comment on the topology of the
TPRM ground-state. One of the main conclusions of our paper is that the TPRM
ground-state is {\it non-separable} at least in the particular geometry we use.
This means that the optimization of the TPRM ground-state can not be done in
two quasi-independent steps.

The structure of the paper is as follows. In section two we formulate the
problem and outline the relevant scaling exponents to be studied later.
Section three discusses the numerical method. In section four we give the
numerical data concerning the scaling behavior. Finally section five finishes
the paper with conclusions.

\section{Two directed polymers in a random medium}

The continuum Hamiltonian for the TPRM problem is written
in all generality as 
\begin{equation}
H = \int_{0}^{t}
\Gamma_1 (\nabla h_1 ({x}))^2 + \Gamma_2 (\nabla h_2 ({x}))^2 
+ V_r (x,h_1) + V_r (x,h_2) + V_{int} dx.
\label{H}
\end{equation} 
The Hamiltonian describes the physics of two elastic lines (subscripts 1 and
2) in the presence of the random potential $V_r$ which is sample-to-sample the
same for both lines. The longitudinal coordinate is labeled with $x$ while the
transverse coordinate (which can be a vector) is $h_1$ or $h_2$. In the
following we shall consider only two 'identical' lines, that is the line
stiffnesses $\Gamma_i$ are taken to be finite and equal. The random potential
$V_r$ describes point disorder and therefore the correlator $\langle V_r (x,h)
V_r(x',h')\rangle \propto \delta(x-x') \delta (h-h')$.
 
The interaction potential $V_r$ gives rise to a variety of phenomena. First,
for ground-state problems the case of a attractive potential is obviously
trivial: the two lines will localize to the same ground-state. In this paper we
are going to deal with a hard-core interaction between the lines 1 and 2. This
implies a delta-function-like $V_r \sim V_0 \delta (x_1 - x_2) \delta (h_1 -
h_2)$ with $V_0 \rightarrow \infty$ so that overlap between the lines is
strictly excluded. Would one allow for e.g. a finite $V_0$ then the one-line
ground-state would act as a pinning defect and the physics would slowly
cross-over from the hard-core case to that of two independent lines as $V_0$
is decreased.

The simplest scaling picture for the TPRM in the presence of a hard-core
interaction $V_r$ consists of two independent directed polymers one being in
the one-line global minimum and the second being in the first local minimum or
the first excited state. This picture implies that the TPRM ground-state would
be {\it separable}, that is it could be constructed by a successive
optimization procedure. This turns out to be {\it false}, but the construction
gives a definition for the effective interaction energy
\begin{equation}
V_{int,eff} = E_1 + E_1^{'} - E_2 \sim L^{\theta_V}
\end{equation}
where $E_1$ refers to the single-DP ground-state energy in a
particular sample, $E_1^{'}$ to the first excited state, $E_2$ is the
true TPRM ground-state energy and $L$ is the system size to be defined
below in section \ref{results}. $\theta_V$ defines a scaling exponent
for this particular form of the interaction energy.  Recall that one
has $E_1\sim A L + \bar{A} L^\theta_1 + \dots$ and that the same is
expected of $E_1^{'}$ as well where $A$, $\bar{A}$ are disorder and
dimension-dependent non-universal pre-factors.  The argument is,
however, essentially based on the claim that in the DPRM problem there
is only one energy scale, that governed by the DPRM energy fluctuation
exponent $\theta$ and is therefore only qualitative. For $E_2$ it is
to be expected that the scaling is of the same form $E_2 \sim B L +
\bar{B} L^{\theta_2}$ where the exponent $\theta_2$ measures the
energy fluctuations of the TPRM ground-state.  The ensemble-averaged
$V_{int,eff}$ allows one to note that since the energy and its
fluctuations have as an upper bound the separable trial ground-state
$\theta_V$ should be limited from above by $\theta_1$.

\begin{figure}[htb]
\narrowtext
\epsfxsize=2.0in 
 ~~~\epsfbox{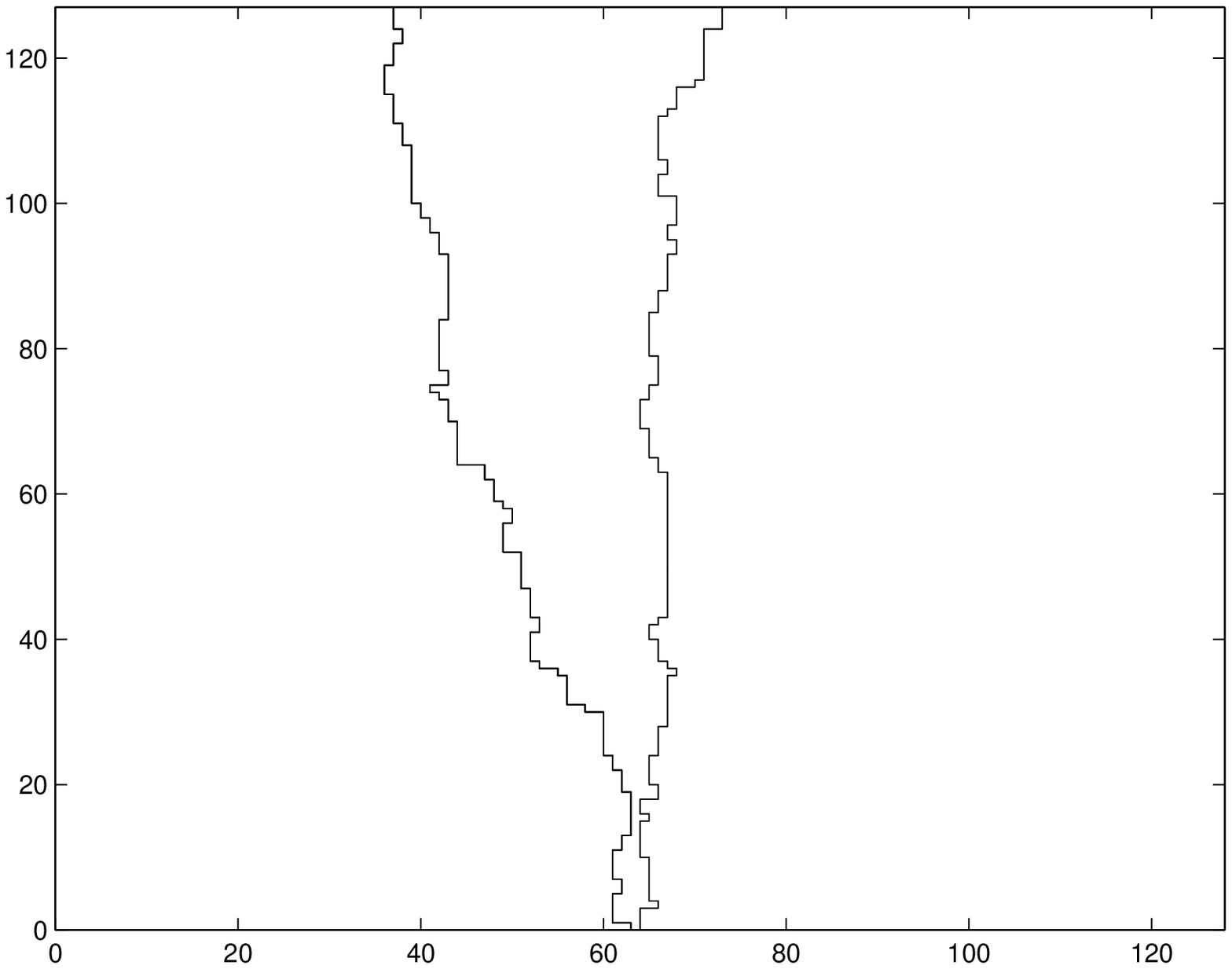}  
\epsfxsize=2.0in 
 ~~~\epsfbox{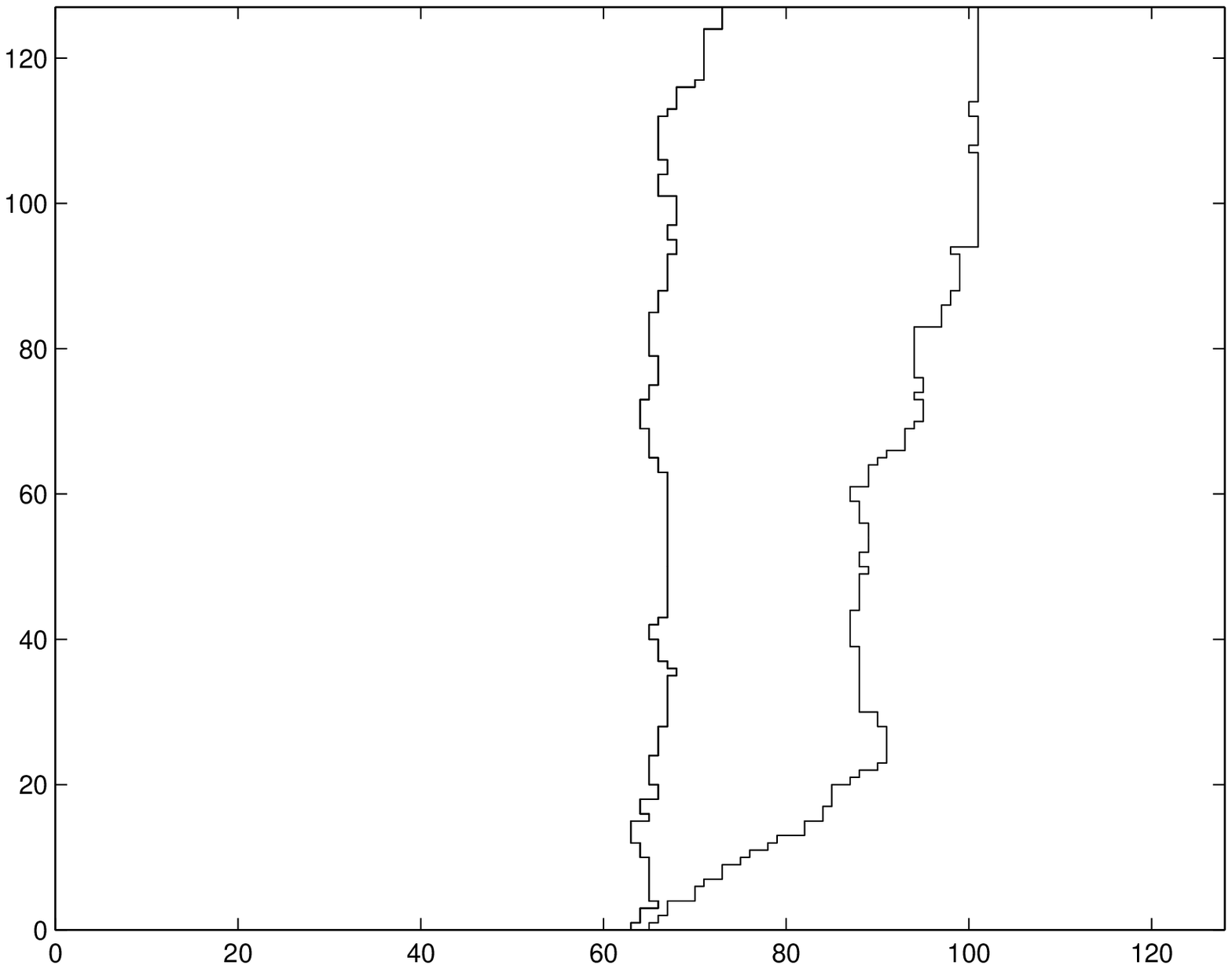}  

\epsfxsize=2.0in 
 ~~~\epsfbox{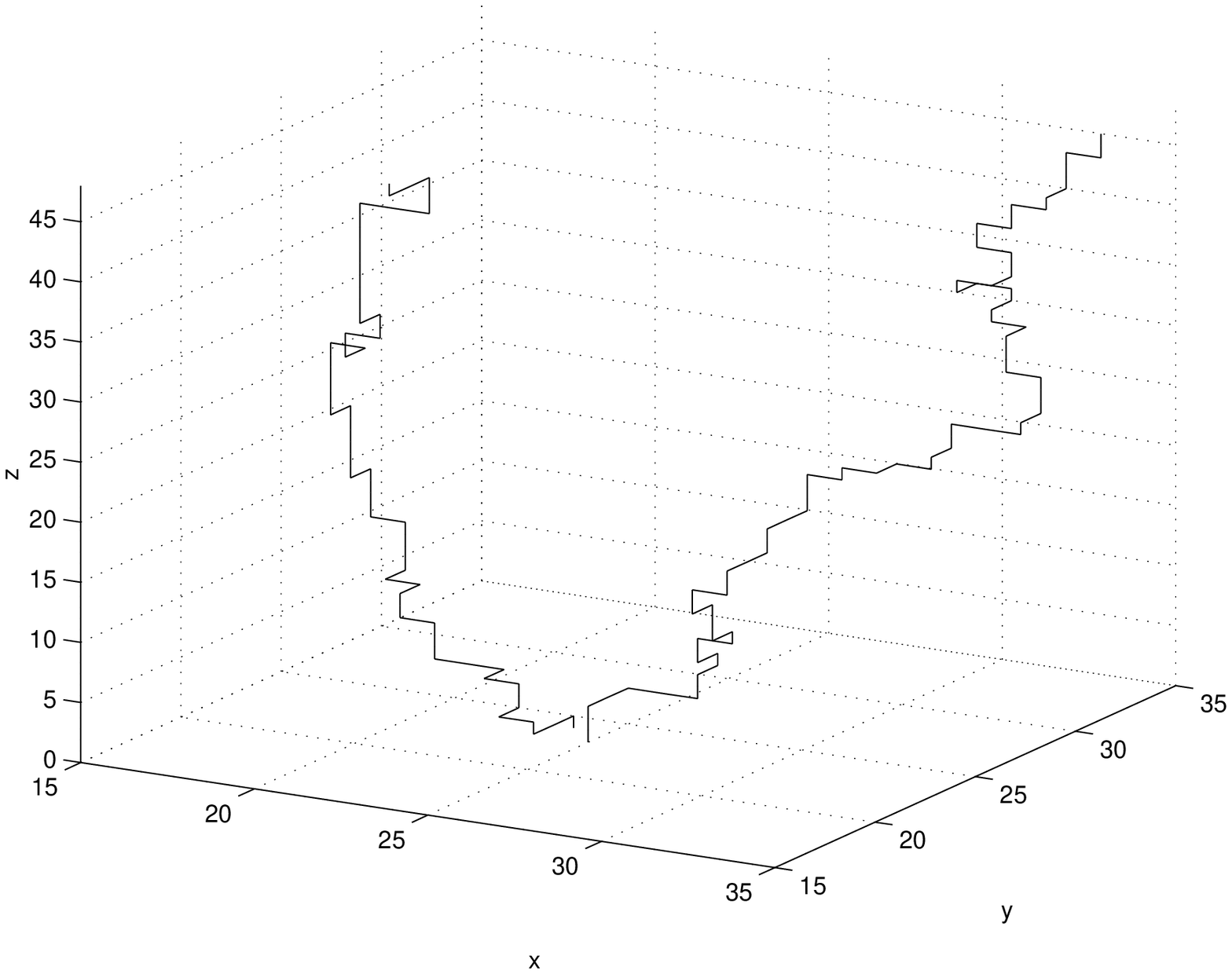}  
\epsfxsize=2.0in 
 ~~~\epsfbox{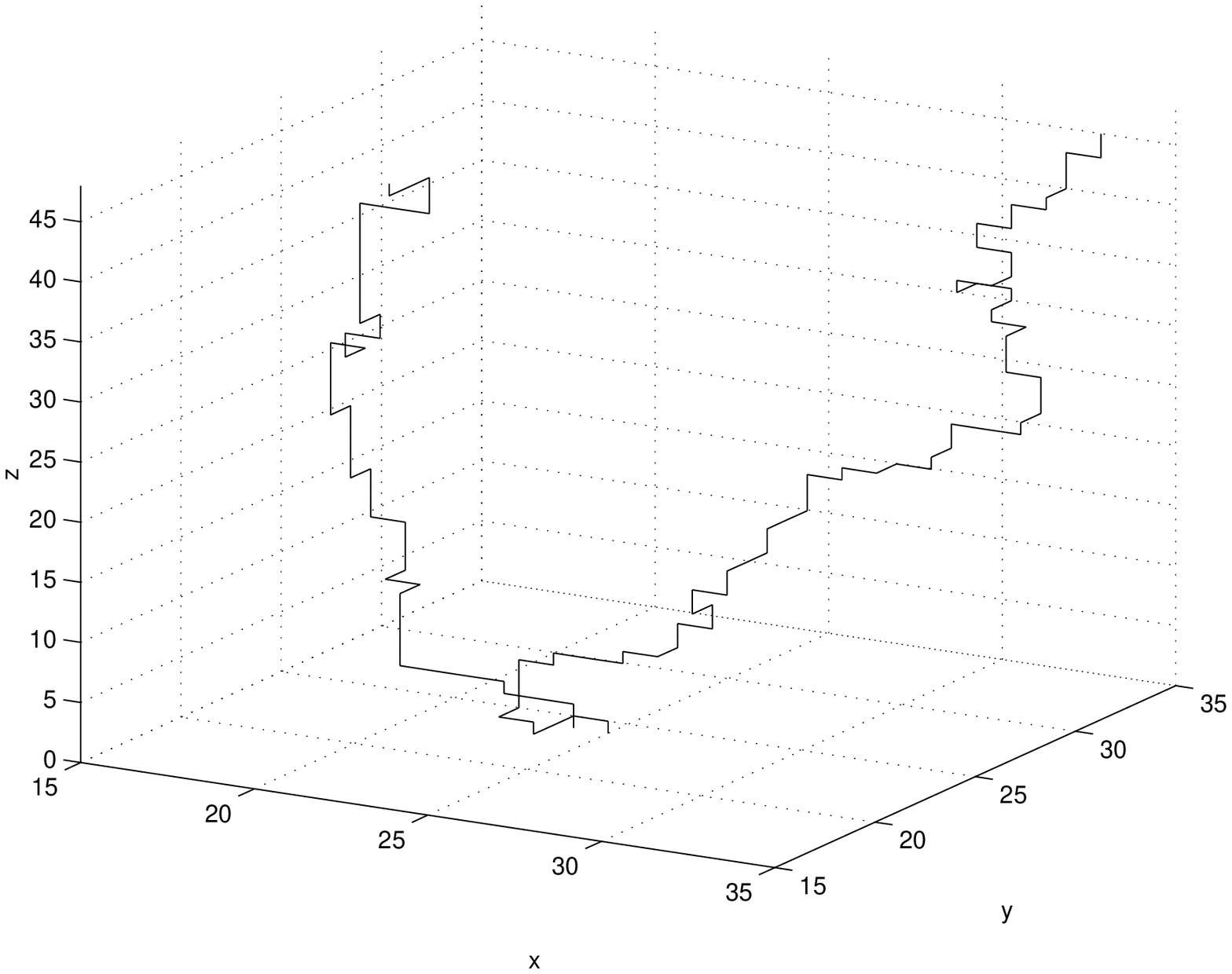}  
\vspace*{5mm}
\caption{
  a): Two-polymer ground-state in 2D, b) the same system but with the first
  (1-line GS) frozen first. c) the TPRM GS in 3D, d) as b) but in 3D. In both
  the 2D and the 3D comparisons the disorder landscape is the same. 
}
\label{example}
\end{figure}

Likewise, the interaction energy can be described by the energy of the TPRM
ground-state minus twice the single line energy, i.e.
\begin{equation}
\delta E_2 = E_2 - 2 E_1 \sim L^{\theta_E}.
\end{equation}
Here $\theta_E$ defines another scaling exponent characterising the
TRPM groundstate.  One has naturally $\delta E_2 + V_{int,eff} =
E_1^{'} - E_1 > 0$ and in particular if the single-line problem has
two geometrically independent, energetically degenerate solutions then
the sum is zero.  Since $\delta E_2$ is positive semi-definite
sample-to-sample, a lower limit for $\theta_2$ is $\theta_1$ and
therefore by this dual construction one would expect that $\theta_2 =
\theta_1$.  In this work we do not consider the roughness properties
of the two-line system but note that for it one would likewise expect
that $\zeta_2 = \zeta_1$. Figure (\ref{example}) shows examples from
two and three dimensions of situations in which the true TPRM
ground-state is {\it non-separable}, i.e. it can not be constructed
out of the states with energies $E_1$ and $E_1^{'}$ and has thus a
non-zero $V_{int,eff}$.

\section{Numerical method}

Here we define the lattice version of the continuum model of two
random polymers with hard core interactions in a random environment
introduced in the preceeding section. We formulate it in such a way
that the connection to a minimum cost flow problem becomes obvious
\cite{heiko,flows}, for which powerfull algorithms from combinatorial
optimization exist that find exact ground states in polynomial time
\cite{review}.

\begin{figure}[htb]
\narrowtext
\epsfysize=.5\columnwidth{\epsfbox{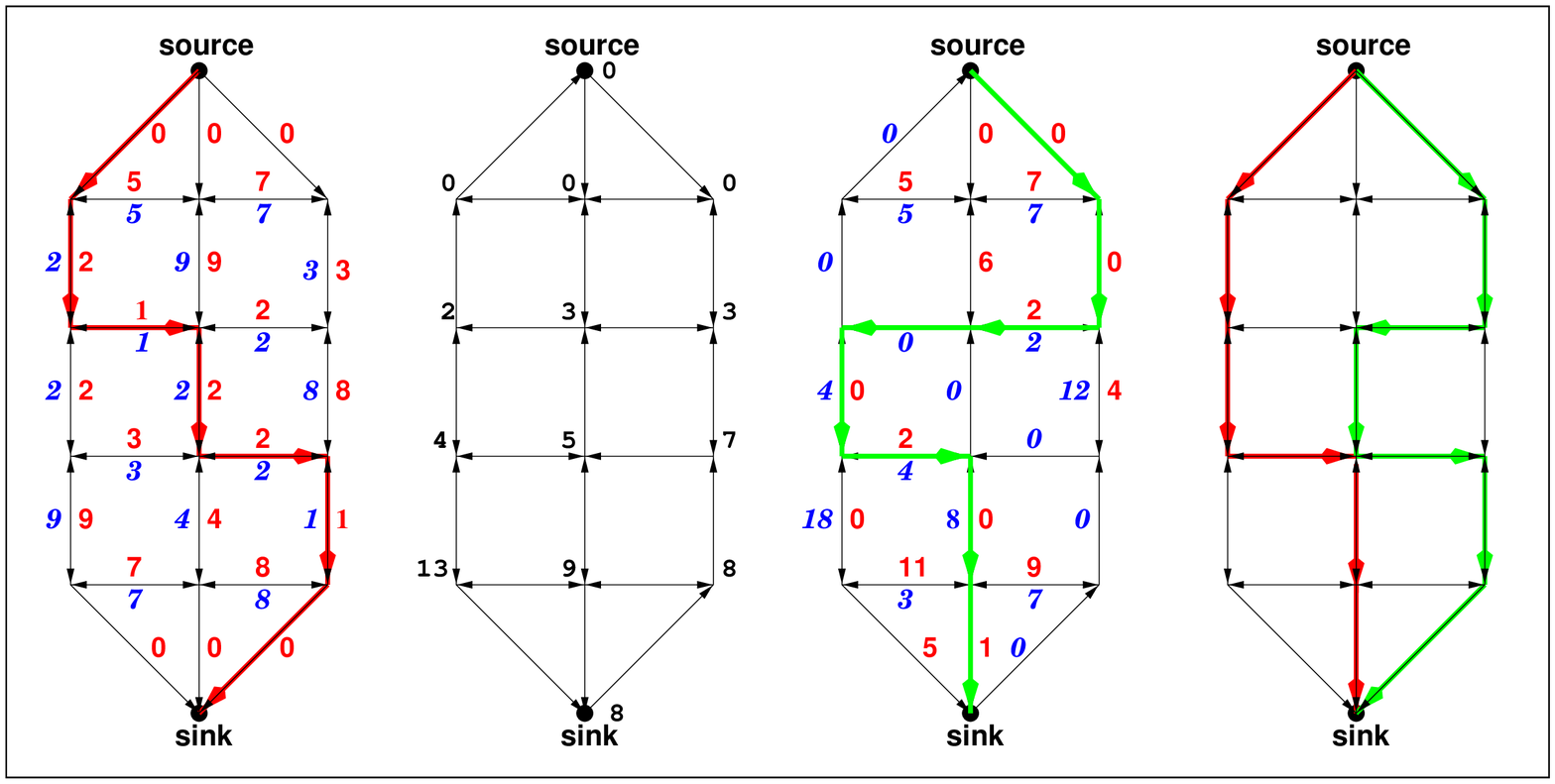}}
\vspace*{12mm}
\caption{
  Sketch of the successive shortest path algorithm for the solution of the
  minimum cost flow problem described in the text.  {\bf (a)} Network for
  $N=0$, the numbers are the reduced costs $e_{ij}$, red for downward and
  right arcs, blue for upward and left arcs.  The red line is a shortest path
  from $s$ to $t$. {\bf (b)} $G_c^0$ with the updated node potentials. {\bf
    (c)} $G_c^0$ with the updated reduced costs. The green line is a shortest
  path. {\bf (d)} OPtimal flow configuration for $N=2$.  }
\label{sketch}
\end{figure}

 Consider the energy function
\be
H({\bf x})=\sum_{(ij)} e_{ij}\cdot x_{ij}\;,
\label{hamilflux}
\ee
where $\sum_{(ij)}$ is a sum over all {\it bonds} $(ij)$ joining site
$i$ and $j$ of a $d$-dimensional lattice, e.g.\ a rectangular
($L^{d-1}\times H$) lattice, with periodic boundary conditions (b.c.)
in $d-1$ space direction and free b.c.\ in one direction.  The bond
energies $e_{ij}\ge0$ are quenched random variables that indicate how
much energy it costs to put a segment of a polymer on a specific bond
$(ij)$. The variables describing the two polymers are
$x_{ij}\in\{0,1\}$ (for hard core interactions), $x_{ij}=1$ if there is
a polymer passing bond $(ij)$ and zero otherwise. For the
configuration to form {\it lines} on each site of the lattice all
incoming flow should balance the outgoing flow, i.e.\ the flow is
divergence free
\be
\nabla\cdot{\bf x}=0\;,
\label{divfree}
\ee
where $\nabla\cdot$ denotes the lattice divergence. Obviously the
flux-line has to enter, and to leave, the system somewhere.  We attach
all sites of one free boundary to an extra site (via energetically
neutral arcs, $e=0$), which we call the source $s$, and the other side
to another extra site, the target, $t$ as indicated in fig.\ 1a. Now
one can push one line through the system by inferring that $s$ has a
source strength of $+1$ and that $t$ has a sink strength of $-1$,
i.e.\
\be
(\nabla\cdot{\bf x})_s=+N\quad{\rm and}\quad
(\nabla\cdot{\bf x})_t=-N\;,
\label{stdiv}
\ee
with $N=1$. Thus, the $1$-line problem consists in minimizing the
energy (\ref{hamilflux}) by finding a flow ${\bf x}$ in the network
(the lattice plus the two extra sites $s$ and $t$) fulfilling the
constraints (\ref{divfree}) and (\ref{stdiv}). Naively one would
expect that the 2-line problem consists simply in adding a second line
to the 1-line configuration, avoiding the bonds already occupied due
to the hard core interaction we consider here. A glance at Fig.
\ref{sketch} convinces us that this is not correct and actually the
main issue of the present paper is to provide evidence that the
correct solution of the TPRM problem is significantly different from
what one gets when assuming the separability of the ground state.

The first key ingredient to treat the two-line problem (and the
$N$-line problem in general \cite{heiko}) is that one does not work
with the original network but with the residual network corresponding
to the actual flux-line configuration, which contains also the
information about possibilities to send flow backwards (now with
energy $-e_{ij}$ since one wins energy by reducing $x_{ij}$), i.e.\ to
modify the actual flow.  Suppose that we put one flux-line along a
shortest path $P(s,t)$ from $s$ to $t$, which means that we set
$x_{ij}=1$ for all arcs on the path $P(s,t)$. Then the residual
network is obtained by reversing all arcs and inverting all energies
along this path, indicating that here we cannot put any further flow
in the forward direction (since we assume hard-core interaction, i.e.\ 
$x_{ij}\le1$), but can send flow backwards by reducing $x_{ij}$ on the
forward arcs by one unit. This procedure is sketched in Figure
\ref{sketch}.

The second key ingredience is the introduction of a so called
potential ${\bf\varphi}$ that fulfills the relation
\be 
\varphi(j)\le \varphi(i)+e_{ij} 
\label{potential}
\ee 
for all arcs $(ij)$ in the residual network, indicating how much
energy $\varphi(j)$ it would {\it at least} take to send one unit of flow
from $s$ to site $j$, IF it would cost an energy $\varphi(i)$ to send it
to site $i$. With the help of these potentials one defines the reduced
costs
\be
c_{ij}^{\bf\varphi}=e_{ij}+\varphi(i)-\varphi(j)\ge0\;.
\label{redcost}
\ee
The last inequality, which follows from the properties of the
potential ${\bf\varphi}$ (\ref{potential}) actually ensures that there
is no loop ${\cal L}$ in the current residual network (corresponding
to a flow {\bf x}) with negative total energy, since
$\sum_{(ij)\in{\cal L}}e_{ij} = \sum_{(ij)\in{\cal
    L}}c_{ij}^{\bf\varphi}$, implying that the flow ${\bf x}$ is
optimal\cite{flows}. 

The idea of the {\it successive shortest path algorithm} is to start
with an empty network, i.e. ${\bf x}^0=0$, which is certainly an
optimal flow for $N=0$, and set ${\bf\varphi}=0$,
$c_{ij}^{\bf\varphi}=e_{ij}$. One now successively adds FL to the
system using the following iteration: Suppose we have an optimal
$N-1$-line configuration corresponding to the flow ${\bf x}^{N-1}$.
The current potential is ${\bf\varphi}^{N-1}$, the reduced costs are
$c_{ij}^{N-1}=e_{ij}+\varphi^{N-1}(i)-\varphi^{N-1}(j)$ and we
consider the residual network $G_c^{N-1}$ corresponding to the flow
${\bf x}^{N-1}$ with the reduced costs $c_{ij}^{N-1}\ge0$. The
iteration leading to an optimal $N$-line configuration $x_{ij}^{N}$ is

\itemsep=0cm
\begin{itemize}
\item[1.] Determine shortest distances $d(i)$ from $s$ to all other
  nodes $i$ with respect to the reduced costs $c_{ij}^{N-1}$ in the
  residual network $G_c^{N-1}$.
\item[2.] For all nodes $i$ update the potential:
  $\varphi^{N}(i)=\varphi^{N-1}(i)+d(i)-d(t)$.
\item[3.] Let $P(s,t)$ denote a shortest path from node $s$ to $t$. To
  obtain $x_{ij}^{N}$ increase (decrease) by one unit the flow
  variables $x_{ij}^{N-1}$ on all forward (backward) arcs $(ij)$ along
  $P(s,t)$.
\end{itemize}

(see Fig. \ref{sketch}). Note that due to the the fact that the
numbers $d(i)$ are shortest distances one has again $c_{ij}^{N}\ge0$,
i.e.\ the flow ${\bf x}^{N}$ is indeed optimal. To estimate the
complexity of this algorithm it is important to note that it is not
necessary to determine shortest paths from $s$ to {\it all} other
nodes in the network; a shortest path from $s$ to $t$ is sufficient if
one updates the potentials in a slightly different way \cite{flows}.
Thus, the complexity of each iteration is the same as that of
Dijkstra's algorithm for finding shortest paths in a network, which is
${\cal O}(M^2)$ for a naive implementation ($M$ is the number of nodes
in the network). We find, however, for the cases we consider
($d$-dimensional lattices) it roughly scales linearly in $M=L^d$. Thus,
for $N$ flux-lines the complexity of this algorithm is ${\cal
  O}(NL^d)$. 

In Figure \ref{example} we show the true ground state configuration for a
specific disorder configuration in 2d and in 3d and compare it with the
one-line ground state plus the first excited state (the latter defined as the
ground state in the network that is left when the bonds occupied by the
one-line ground state are excluded). This is a typical example in which the
two two-line configuration in 2d and in 3d are distinct.

\section{Results}
\label{results}

For the actual computations reported in the following we set the height of the
system $H$ equal to its lateral size $L$, i.e.\ $H=L$, yielding a square
geometry in 2d and a cubic one in 3d) and considered system sizes from $L=16$
to $L=256$ in 2d and from $L=8$ to $L=64$ in 3d. For each system size the
results are averaged over $N=12000$ (2D) and $N=8000$ (3D) disorder
configurations, and quantities like ${\cal O}=E_1$, $E_2$, $\delta E_2$,
$V_{int,eff}$ denote disorder averages from now on.

We expect the various exponents that we estimate to be independent of the
actual disorder we put in (as long it is uncorrelated and does not have
algebraic tails), nevertheless we took two different probability distributions
for the bond energies: 1) a {\it uniform} distribution for which $P(e_{ij})=1$
for $e_{ij}\in[0,1]$ and 0 otherwise; 2) a {\it binary} distribution in which
$e_{ij}$ is 1 with probability $p$ and 0 with probability $1-p$.

\subsection{Two dimensions}

\begin{figure}[htb]
\narrowtext
\epsfysize=.5\columnwidth{\rotate[r]{\epsfbox{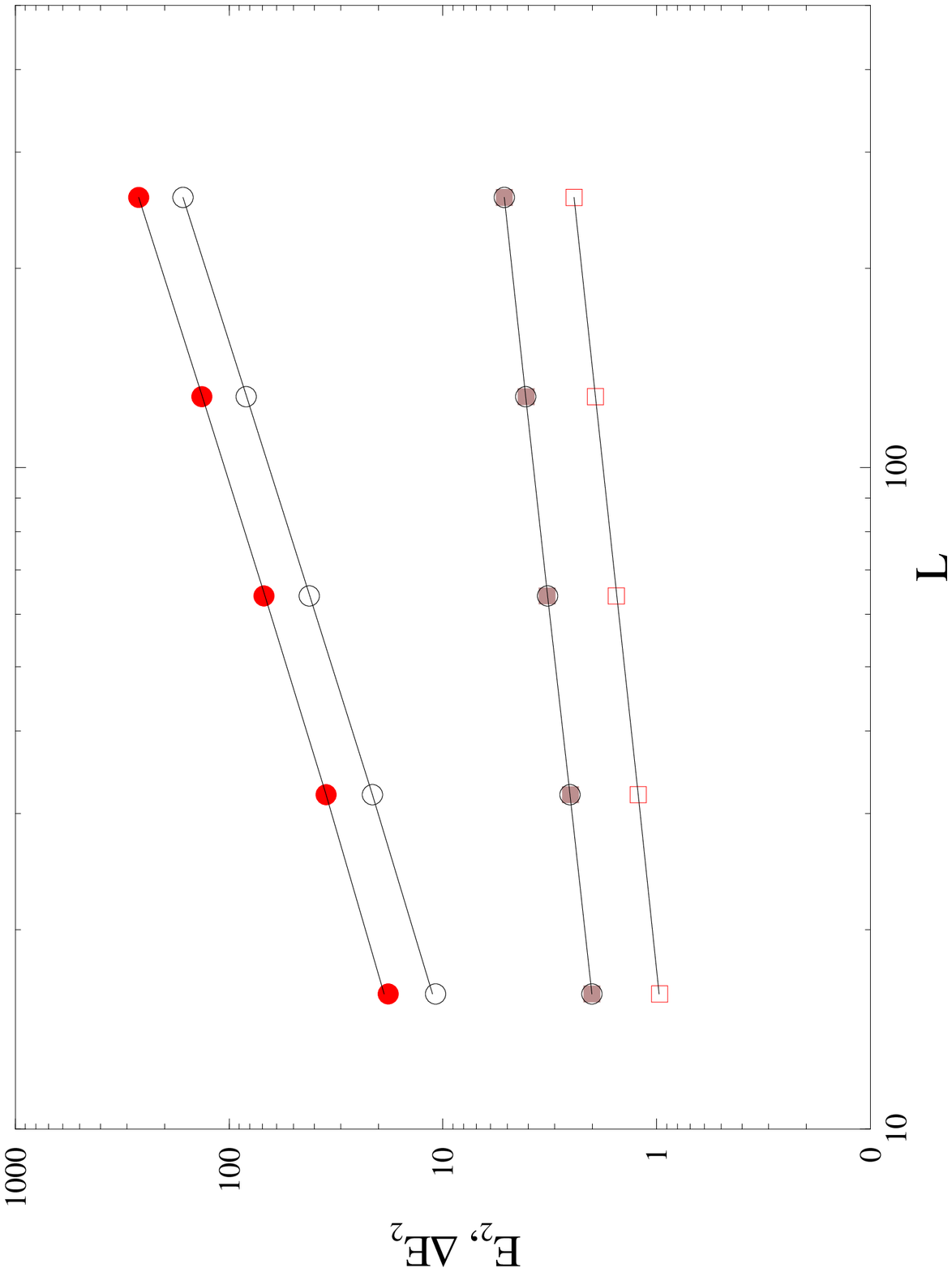}}}
\vspace*{12mm}
\caption{
  Energy $E_2$ (circles) and energy fluctuations $\Delta E_2$ (squares) of the
  TPRM problem in two dimensions in a log-log plot. We show data for binary
  disorder ($e_{ij}\in\{0,1\}$) (filled symbols) and the uniform distribution
  of $e_{ij}$'s (open symbols).  One expects $E_2\propto L$ and $\Delta
  E_2\propto L^\theta$, correspondingly the straight lines have slopes $1$
  (top) and $1/3$ (bottom).  }
\label{2dene}
\end{figure}

Figure \ref{2dene} shows the scaling of the two-line system energy and energy
fluctuations for both a uniform distribution for the $e_{ij}$'s and a binary
one with q$p=0.8$. As expected, the scaling of the total energy $E_2$ is
linear and the fluctuations $\delta E_2$ scale with an exponent $\theta_2$
with $\theta_2 \simeq \theta_1$, the one-line energy fluctuation exponent.
This adheres to the picture that the energetics of the DP problem are in
general dictated by the one-line exponent.

\begin{figure}[htb]
\narrowtext
\epsfysize=.5\columnwidth{\rotate[r]{\epsfbox{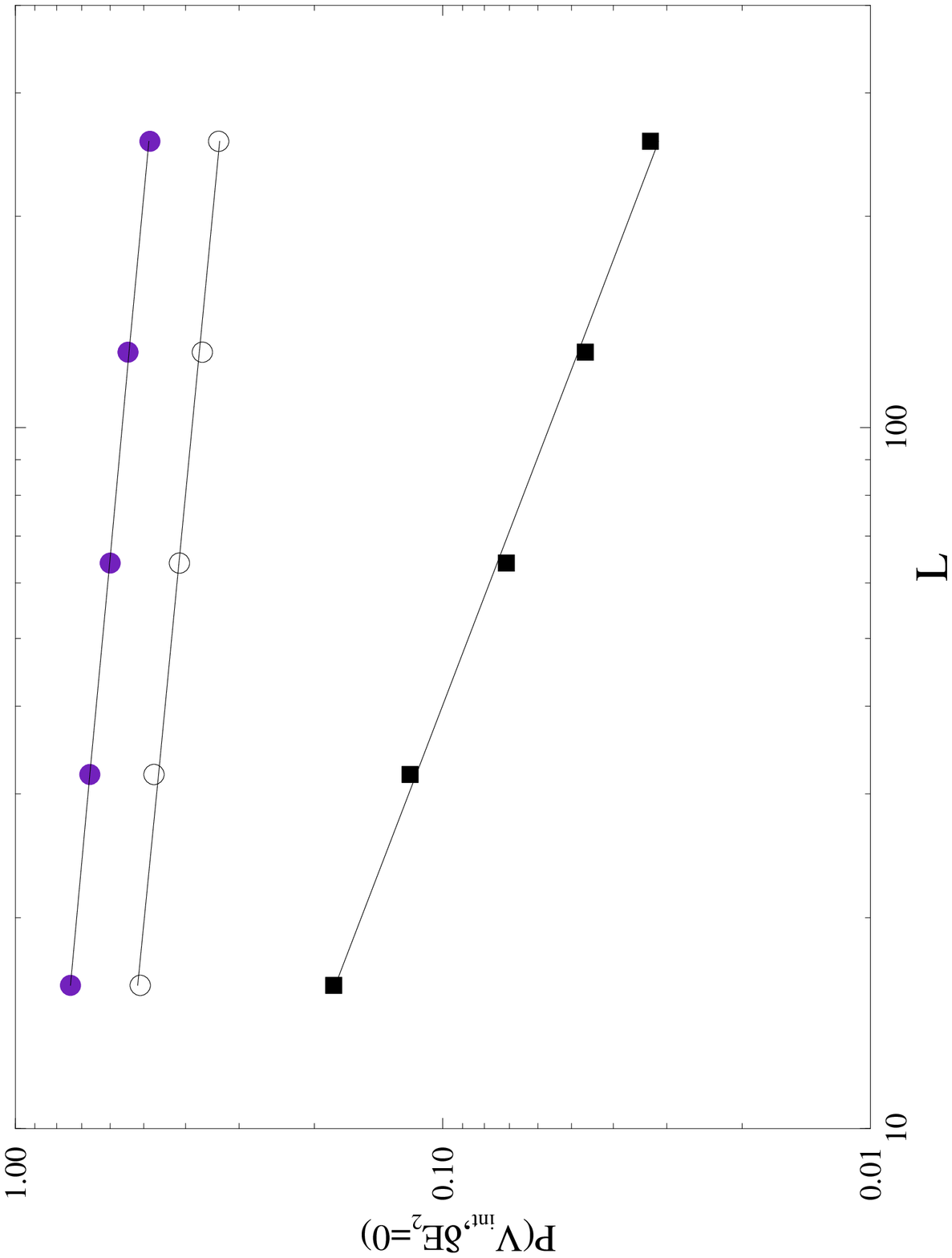}}}
\vspace*{12mm}
\caption{
  $P(\delta E_2=0)$ (squares) and $P(V_{int,eff})=0$ (circles) vs. $L$ in 2d in
  a log-log plot. Data for both binary (filled symbols) and uniform (open
  symbols) distribution of the bond energies $e_{ij}$. The data follow the
  relations $P(\delta E_2=0)\propto L^{-a_1}$ and $P(V_{int,eff})=0\propto
  L^{-a_2}$ with $a_1$ and $a_2$ given by the slopes of the straight lines:
  $a_1=0.63$ ($\approx1-\theta$) and $a_2=0.15$.
}
\label{bpe1}
\end{figure}

In Figure \ref{bpe1} we show the probability that $\delta E_2=0$ as a function
of system size. This measures the true degeneracy of the two-line system as
the joint ground-state can be obtained from two independent minima with the
same energy.  $P(\delta E_2=0) \sim L^{-a_1}$ with $a_1=0.63\pm0.03$ 
which is compatible with to $a_1=2/3$ adhering thus to
Tang's result \cite{tang} which indicated  $a_1 = 1-\theta$.  
One can compare this with the scaling of $P(V_{int,eff}=0)$, which
scales with an exponent $a_2=0.15\pm0.02$ for both
distributions ($P \sim L^{-a_2}$).  Similarly to Tang's conjecture,
we are left with a picture which explains
 the frequency of separable ground-states (with $\delta E_2=0$) by a
picture in which the two lines belong to two neighboring trees in the energy
landscape. This means that one considers an inverted structure in
which the two lines end up next to each other but belonging to
two different trees (starting from $x=L$) with the same energy.
Meanwhile the interaction
energy in general shows increasing entanglement with a probability for a
separable GS that decays with a novel exponent $a_2=0.15$.

\begin{figure}[htb]
\narrowtext
\epsfysize=.5\columnwidth{\rotate[r]{\epsfbox{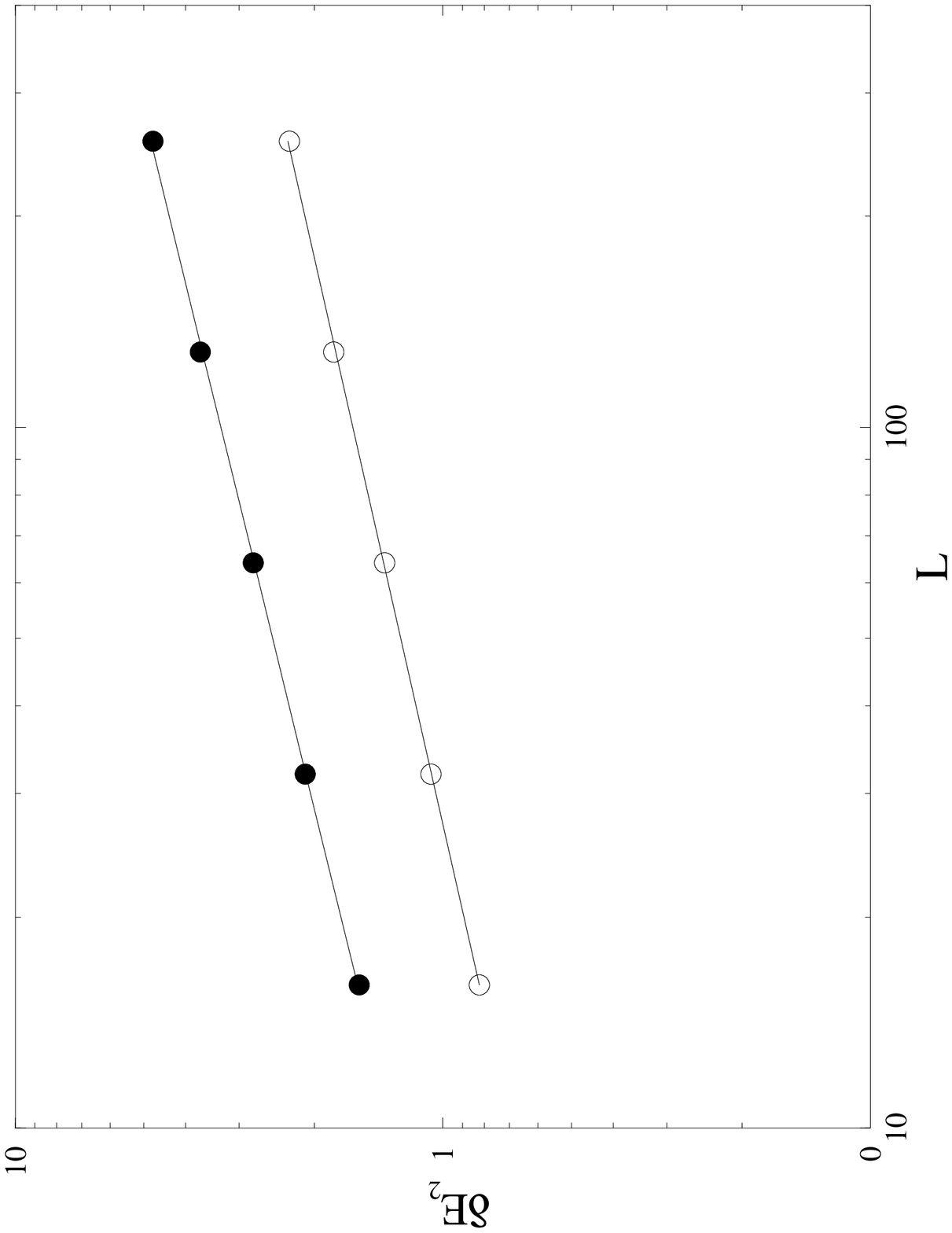}}}
\vspace*{12mm}
\caption{
  $\delta E_2$ in 2d for binary (filled symbols) and uniform (open symbols)
  distribution of the bond energies $e_{ij}$ in a log-log plot. It is $\delta
  E_2\propto L^{\theta_E}$ with $\theta_E = 0.39 \pm 0.02$.
}
\label{intene1}
\end{figure}

\begin{figure}[htb]
\narrowtext
\epsfysize=.5\columnwidth{\rotate[r]{\epsfbox{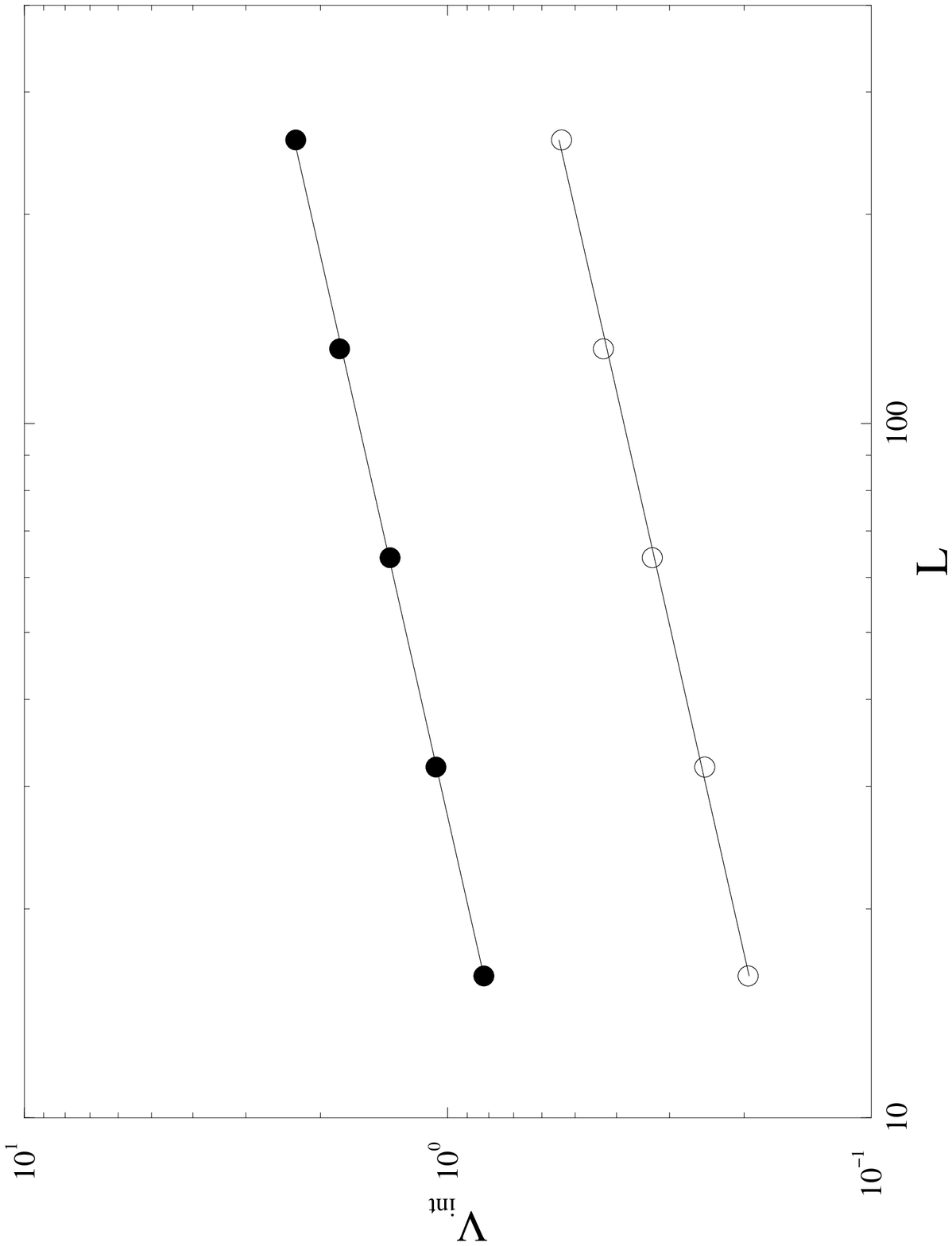}}}
\caption{
  $V_{int,eff}$ in 2d for binary (filled symbols) and uniform (open symbols)
  distribution of the bond energies $e_{ij}$ in a log-log plot. It is
  $V_{int,eff}\propto L^{\theta_V}$ with $\theta_V 0.39 \pm 0.02$.
}
\label{intene2}
\end{figure}

\begin{figure}[htb]
\narrowtext
\mbox{
\epsfysize=.5\columnwidth{\rotate[r]{\epsfbox{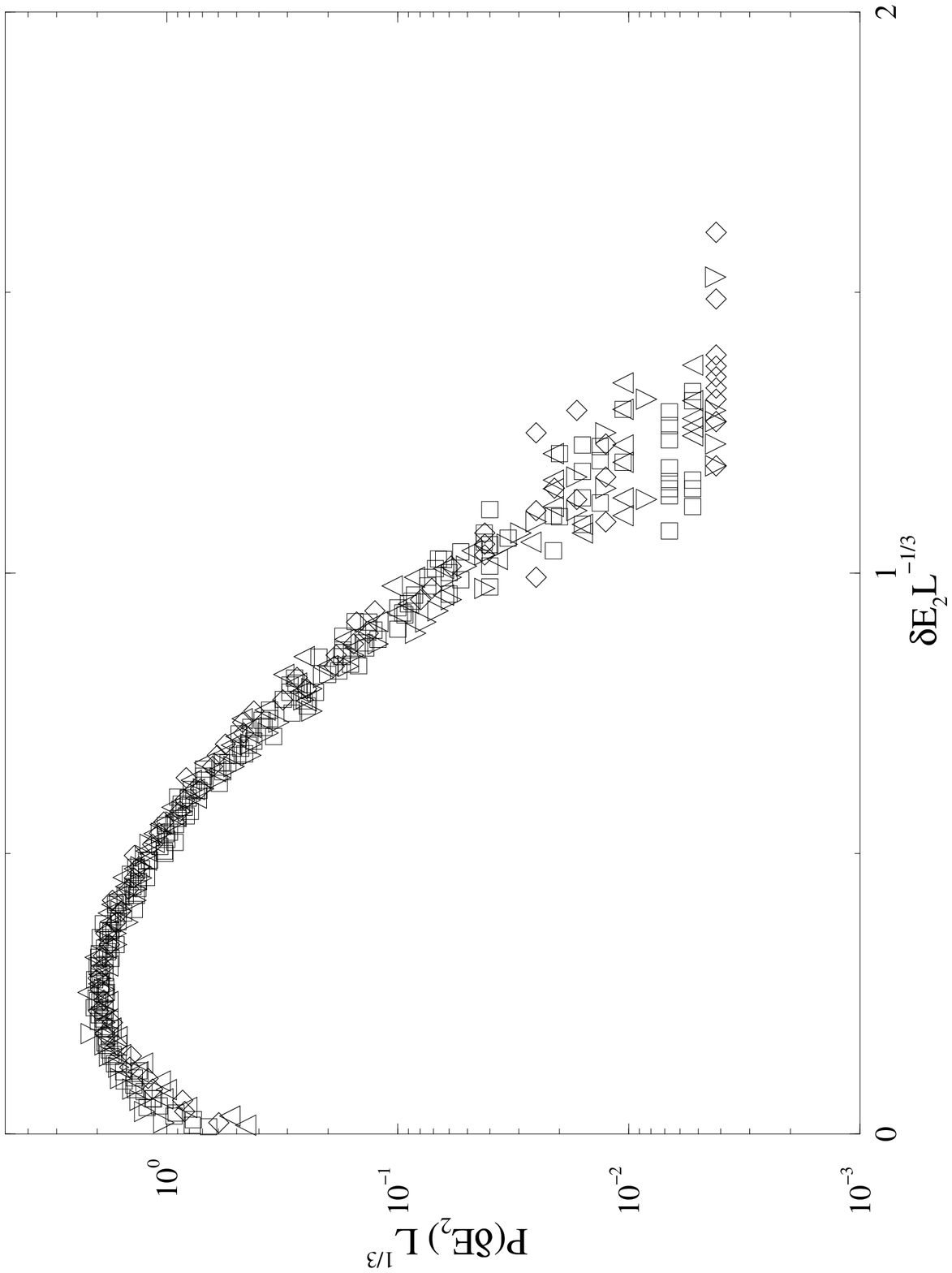}}}
\epsfysize=.5\columnwidth{\rotate[r]{\epsfbox{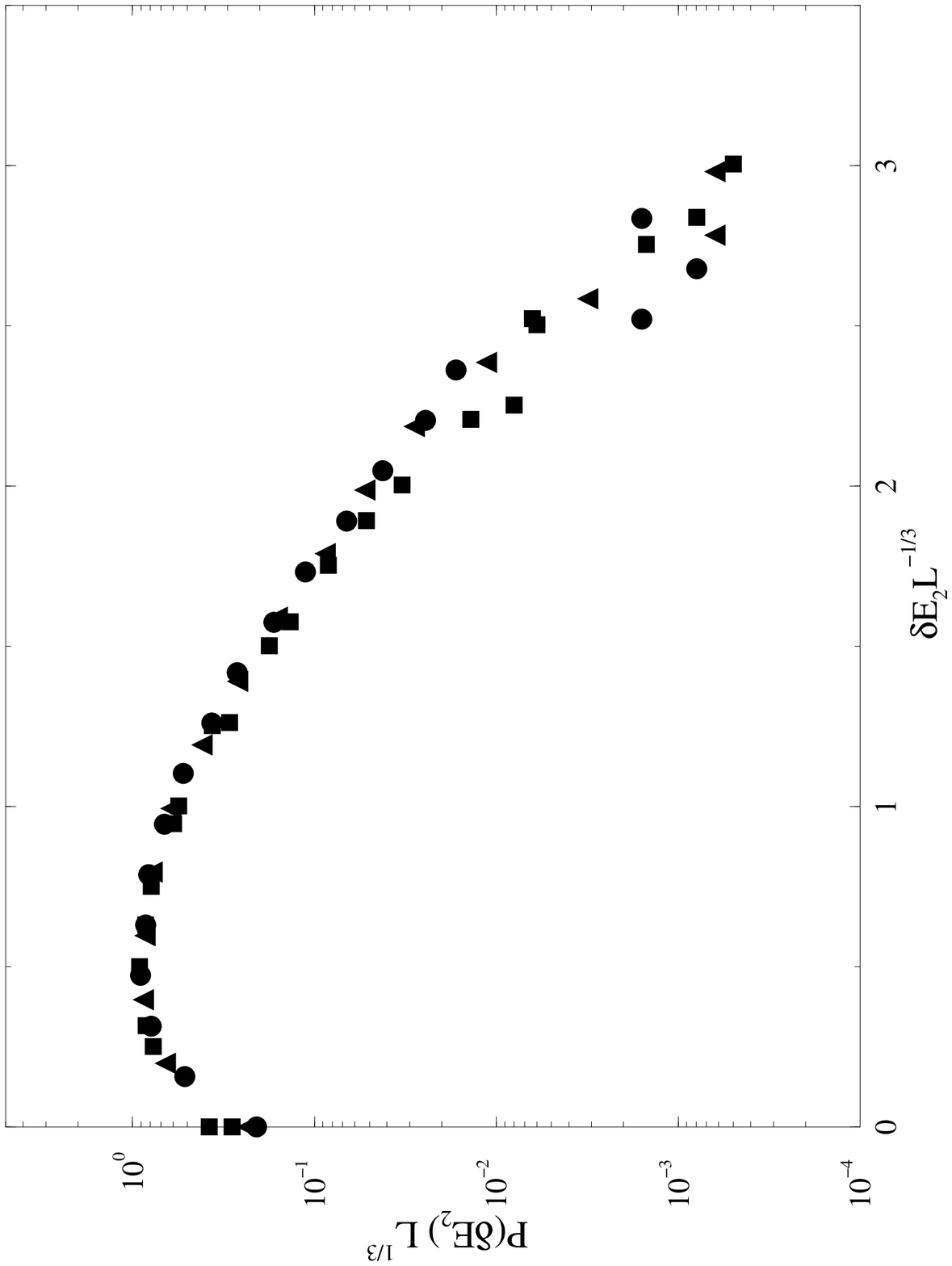}}}
}
\mbox{
\epsfysize=.5\columnwidth{\rotate[r]{\epsfbox{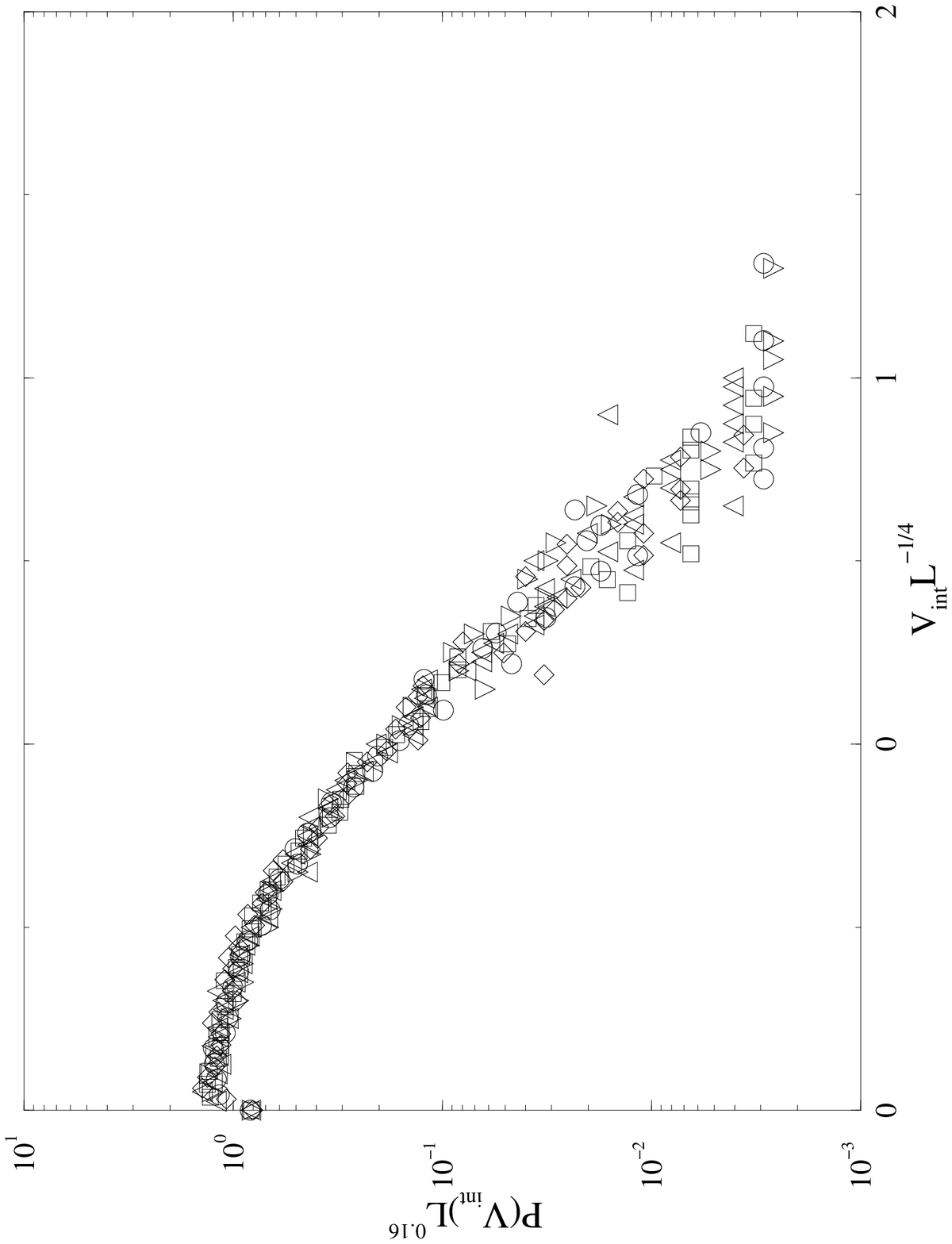}}}
\epsfysize=.5\columnwidth{\rotate[r]{\epsfbox{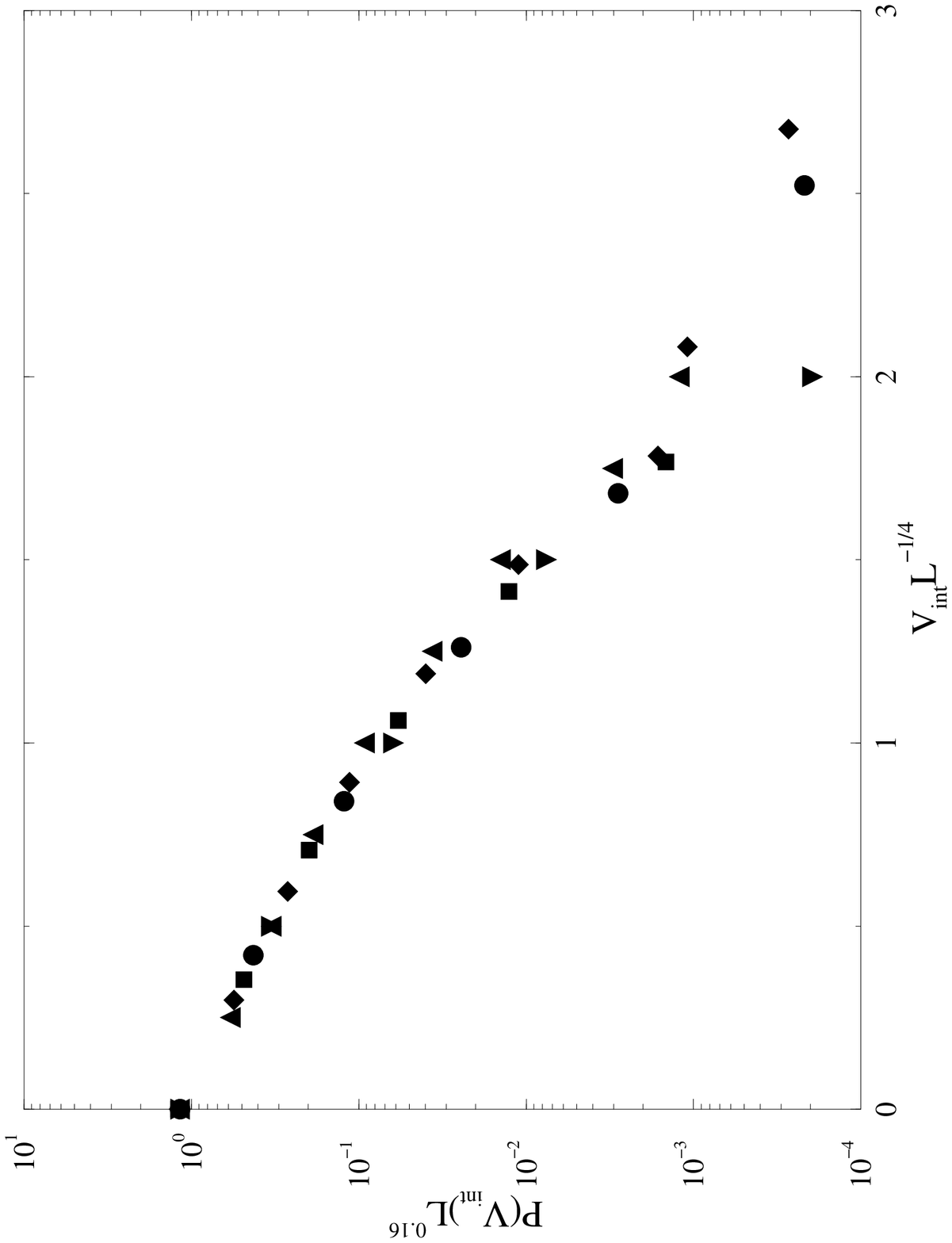}}}
}
\caption{
  Scaling plots of the probability distributions of $\delta E_2$ and
  $V_{int,eff}$ in 2d for binary (filled symbols) and uniform (open symbols)
  distribution of the bond energies $e_{ij}$ in a log-log plot. 
}
\label{collapse}
\end{figure}

Figures \ref{intene1} and \ref{intene2} discuss further the scaling of the
mean interaction energies $\delta E_2$ and $V_{int,eff}$ for the both
distributions. We find the exponents $\theta_E \sim 0.39 \pm 0.03$ and
$\theta_V \sim 0.39 \pm 0.03$, respectively.
For both these quantities we seem to obtain that the effective
scaling exponents are slightly higher than the one- or two-line energy
fluctuation exponents as such. However, as shown in figure \ref{collapse} we
can collapse the energy probablity distributions for $\delta E_2$ and
$V_{int,eff}$ by using a two-exponent collapse. Note that this
is different from the simple collapse using $\theta_E$ and $\theta_V$,
however the two exponents combined make it so that the averages
scale with $\theta_E$ and $\theta_V$.

\subsection{Three dimensions}

\begin{figure}[htb]
\narrowtext
\epsfysize=.5\columnwidth{\rotate[r]{\epsfbox{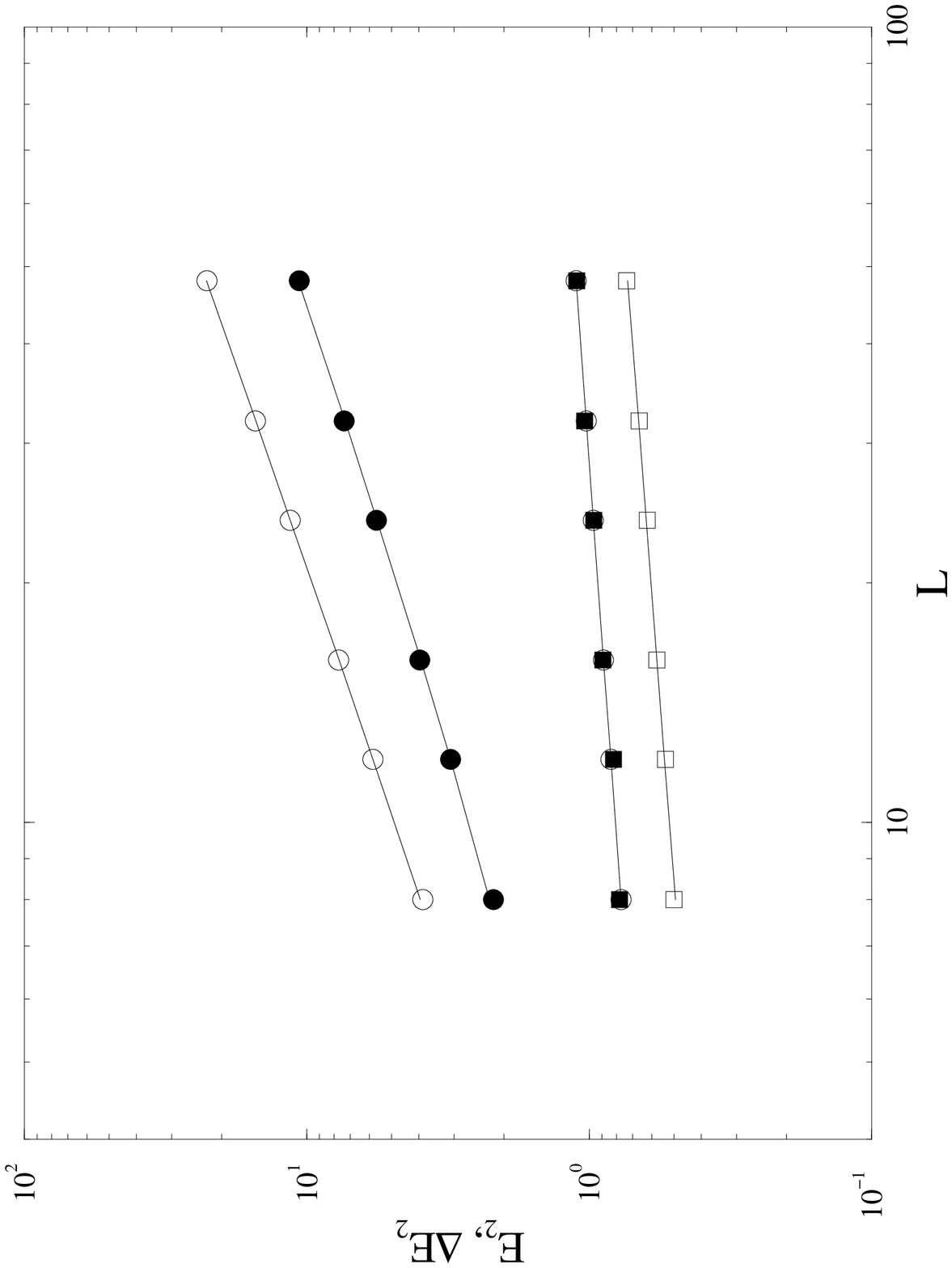}}}
\vspace*{12mm}
\caption{
  Energy $E_2$ (circles) and energy fluctuations $\Delta E_2$ (squares) of the
  TPRM problem in three dimensions in a log-log plot. We show data for binary
  disorder ($e_{ij}\in\{0,1\}$) (filled symbols) and the uniform distribution
  of $e_{ij}$'s (open symbols).  One expects $E_2\propto L$ and $\Delta
  E_2\propto L^\theta$, correspondingly the straight lines have slopes $1$
  (top) and $0.24$ (bottom).
}
\label{3dene}
\end{figure}

Figure \ref{3dene} shows the scaling of three-dimensional case again for both
a uniform distribution for the $e_{ij}$'s and a binary one with $p=0.8$ for
the case of the two-line system energy and energy fluctuations. As expected,
the scaling of the total energy $E_2$ is linear and the fluctuations $\delta
E_2$ scale with an exponent $\theta_2$ with $\theta_2 \simeq
\theta_1\approx0.24$, the one-line energy fluctuation exponent in three
dimensions.

\begin{figure}[htb]
\narrowtext
\epsfysize=.5\columnwidth{\rotate[r]{\epsfbox{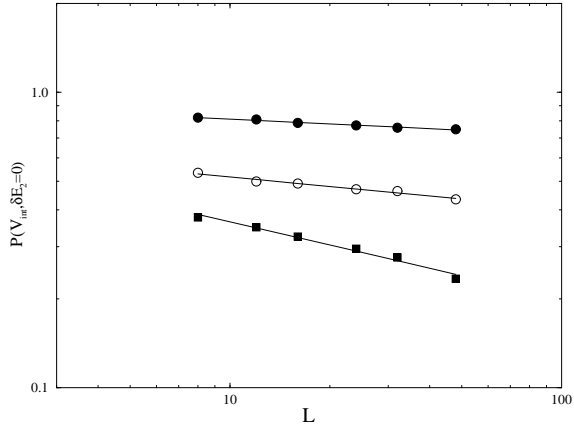}}}
\vspace*{12mm}
\caption{$P(\delta E_2=0)$ (squares) and $P(V_{int,eff})=0$ (circles) vs. $L$ 
  in 3d in a log-log plot. Data for both binary (filled symbols) and uniform
  (open symbols) distribution of the bond energies $e_{ij}$. The data follow
  the relations $P(\delta E_2=0)\propto L^{-a_1}$ and $P(V_{int,eff})=0\propto
  L^{-a_2}$ with $a_1$ and $a_2$ given by the slopes of the straight lines:
  $a_1=0.25$ and $a_2=0.11$, $0.05$ for binary and continuous disorder,
  respectively.
}
\label{bp2e1}
\end{figure}

In Figure \ref{bp2e1} we show the probability that $\delta E_2=0$ as a
function of system size. In 3D, for binary disorder,
$P(\delta E_2=0) \sim L^{-a_1}$ with $a_1 =
0.25 \sim \theta_1$ in contrast with the geometric picture valid in 2D.  The
scaling of $P(V_{int,eff}=0)$ can not be described with a unique exponent
and we find $a_2=0.11\pm0.01$ for binary, and $a_2 = 0.05\pm 0.01$
for continuous disorder ($P \sim L^{-a_2}$).  Again the interaction energy in
general shows increasing entanglement with a probability for a separable GS
that decays with novel exponents $a_1$, $a_2$.

\begin{figure}[htb]
\narrowtext
\epsfysize=.5\columnwidth{\rotate[r]{\epsfbox{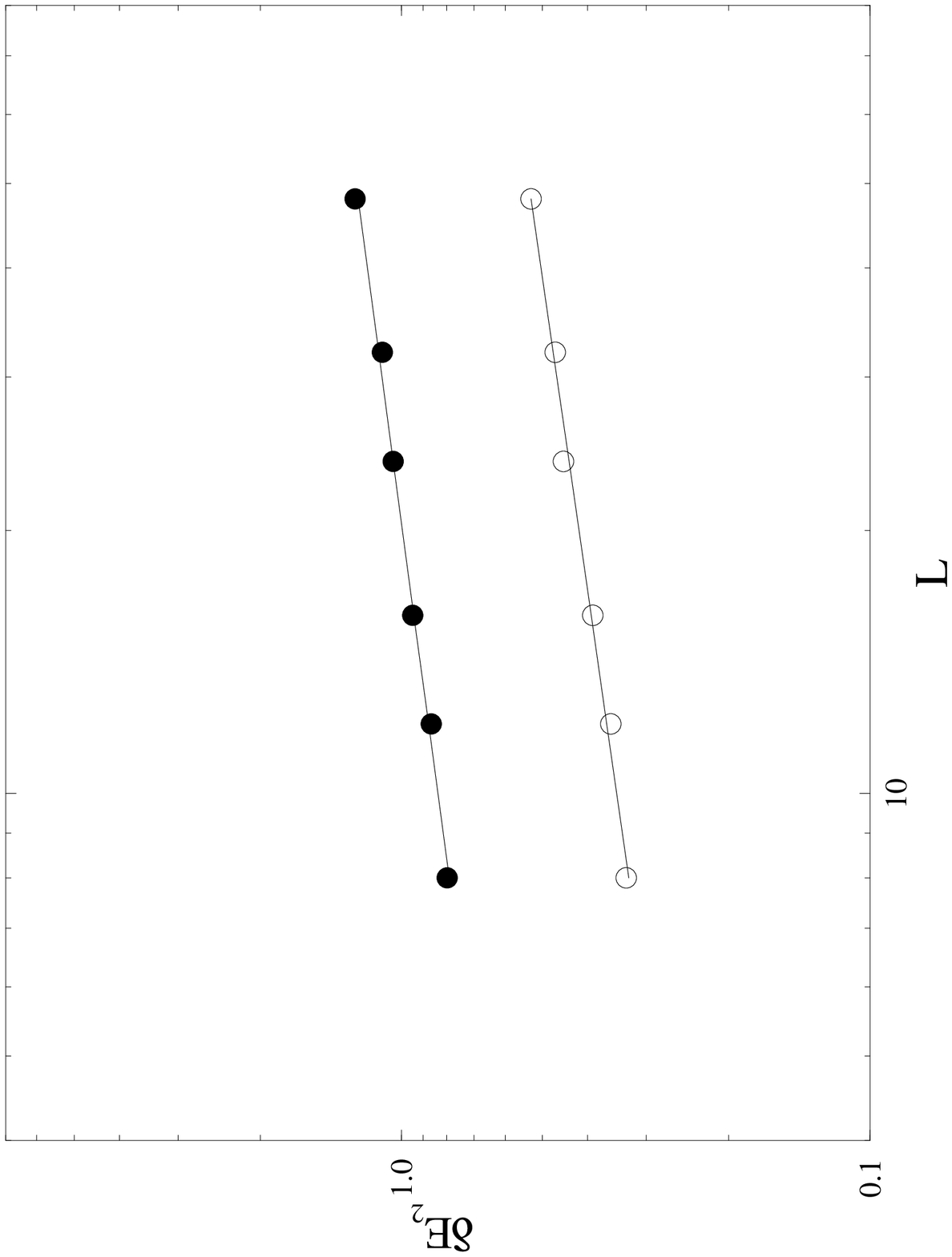}}}
\vspace*{12mm}
\caption{
  $\delta E_2$ in 3d for binary (filled symbols) and uniform (open symbols)
  distribution of the bond energies $e_{ij}$ in a log-log plot.  It is $\delta
  E_2\propto L^{\theta_E}$ with $\theta_E=0.26 \pm 0.02$ 
}
\label{int2ene1}
\end{figure}

\begin{figure}[htb]
\narrowtext
\epsfysize=.5\columnwidth{\rotate[r]{\epsfbox{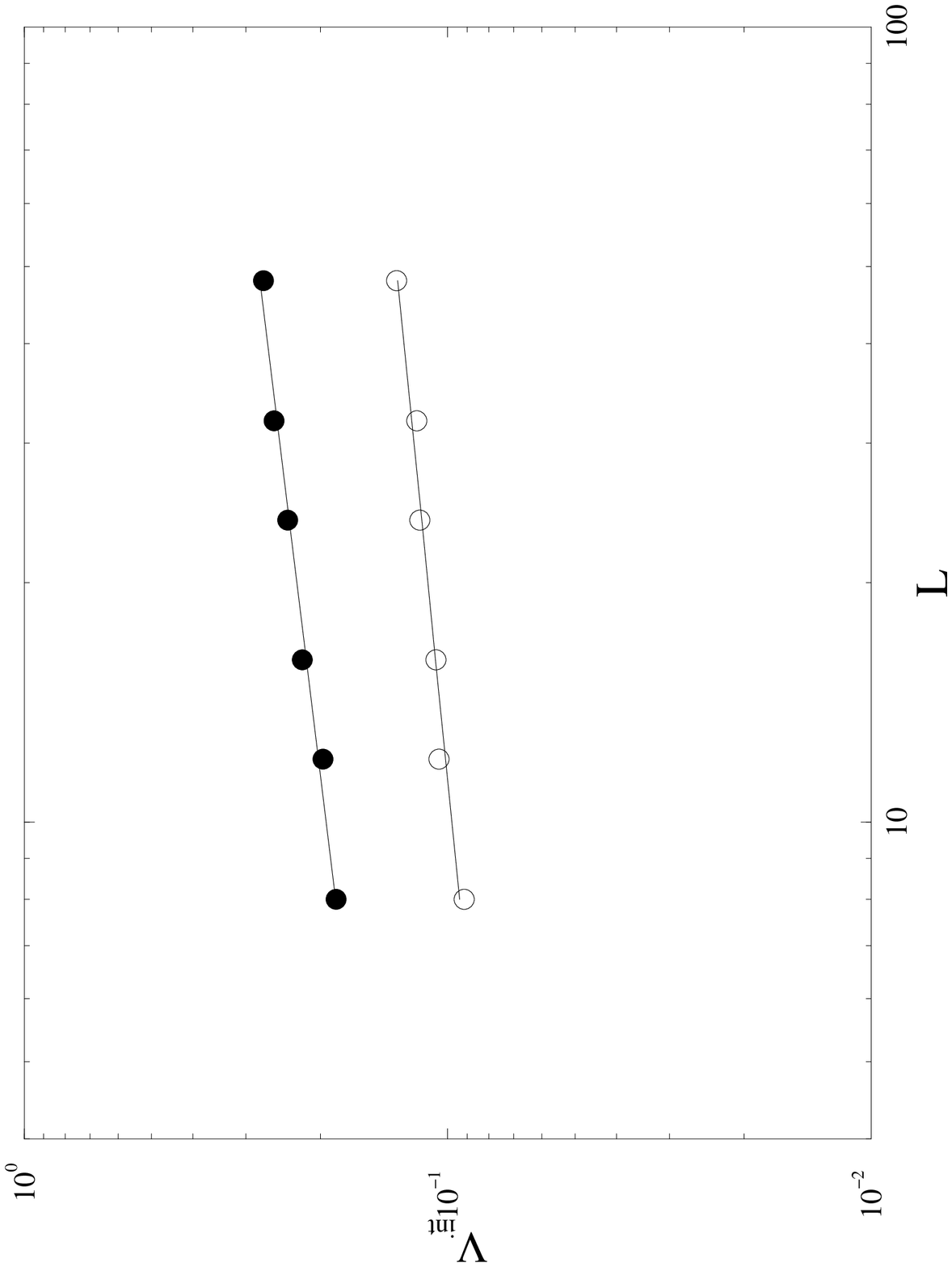}}}
\vspace*{12mm}
\caption{
  $V_{int,eff}$ in 3d for binary (filled symbols) and uniform (open symbols)
  distribution of the bond energies $e_{ij}$ in a log-log plot. It is
  $V_{int,eff}\propto L^{\theta_V}$ with $\theta_V=0.21  \pm 0.02$.
}
\label{int2ene2}
\end{figure}

Figures \ref{int2ene1} and \ref{int2ene2} discuss further the scaling of the
mean interaction energies $\delta E_2$ and $V_{int,eff}$ for the continous
distributions.  The 3D exponents become $\theta_E = 0.26 \pm 0.02$
and $\theta_V = 0.21 \pm 0.02$.
As shown in figure \ref{collapse} we can collapse the energy
probablity distributions for $\delta E_2$ and $V_{int,eff}$ by using as
in 2D a two-exponent collapse. For binary disorder the collapse of the
data makes sense in both cases, for continuous we restrict ourselves
to $\delta E_2$.

\section{Conclusions}
In this paper we have investigated the joint ground-state of two directed
polymers in a random medium, the TPRM problem. The main questions addressed
here are whether the scaling of the TPRM can be described with the one-line
exponents and an associated picture of behavior and if not so when.
Unsurprisingly it turns out that $V_{int,eff}$ as defined here seems to result
in an {\it independent} exponent that can not be explained by the one-line
scaling arguments.  This is natural since it measures the difference of the
true TPRM ground-state to the 'Ansatz' of two separable states and is thus the
first non-analytic and non-trivial correction characterizing the unique nature
of the TPRM problem. On the other hand some of the features of the TPRM
energetics, like the degeneracy of $\delta E_2$ are clearly related to the
single-line picture in two dimensions. In three dimensions this is no longer
true. We lack a geometrical explanation for the scaling of the degeneracy
exponent $a_2$ in this higher-dimensional case.

\acknowledgements This work has been financially supported by the Finnish
Academy of Science (FAS) and the German Academic Exchange Service (DAAD)
within a common exchange project. V.P. would like to acknowledge
the hospitality of University of Cologne.



\vfill
\eject

\vfill
\eject

\begin{figure}[htb]
\narrowtext
\mbox{\epsfysize=.5\columnwidth{\rotate[r]{\epsfbox{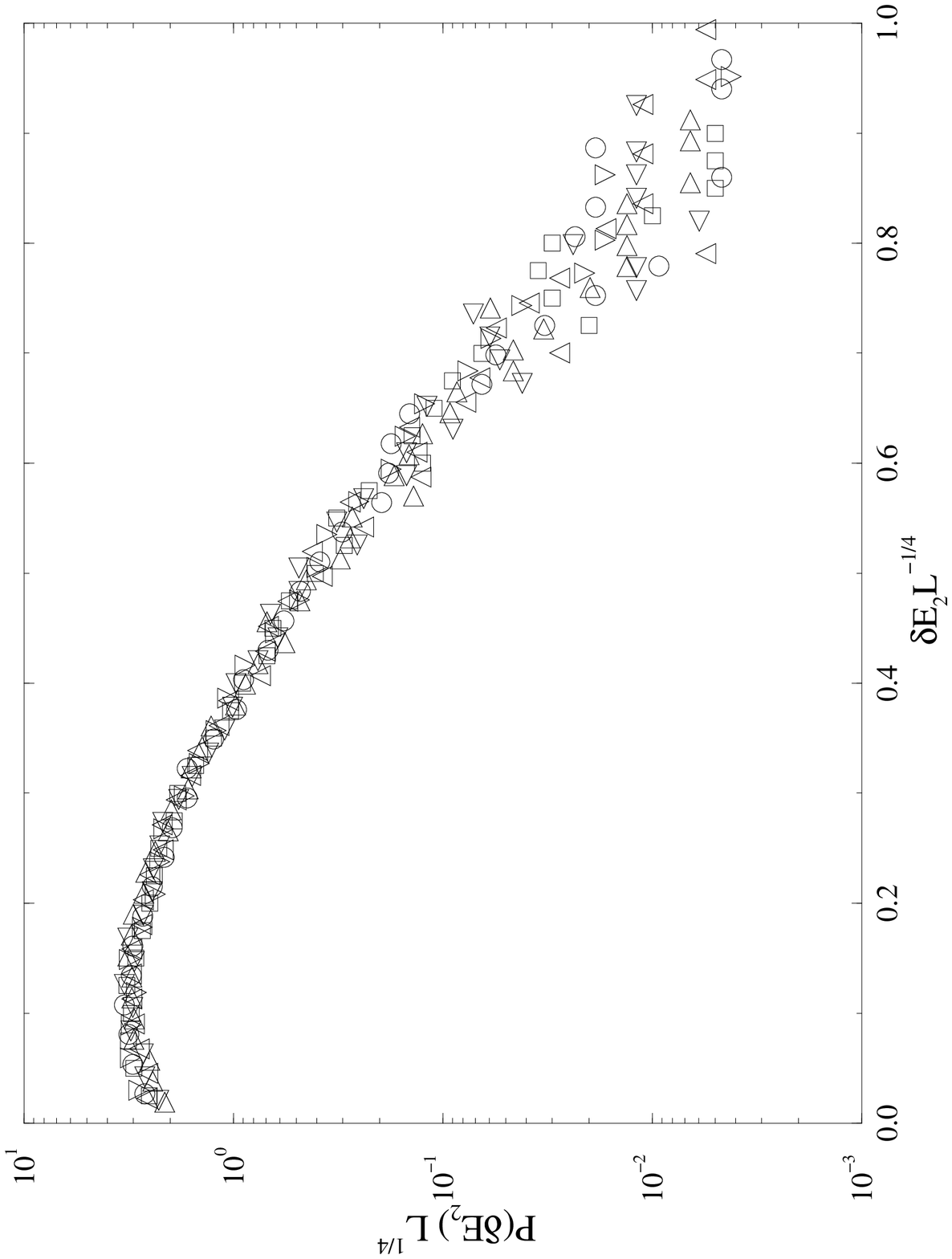}}}
\epsfysize=.5\columnwidth{\rotate[r]{\epsfbox{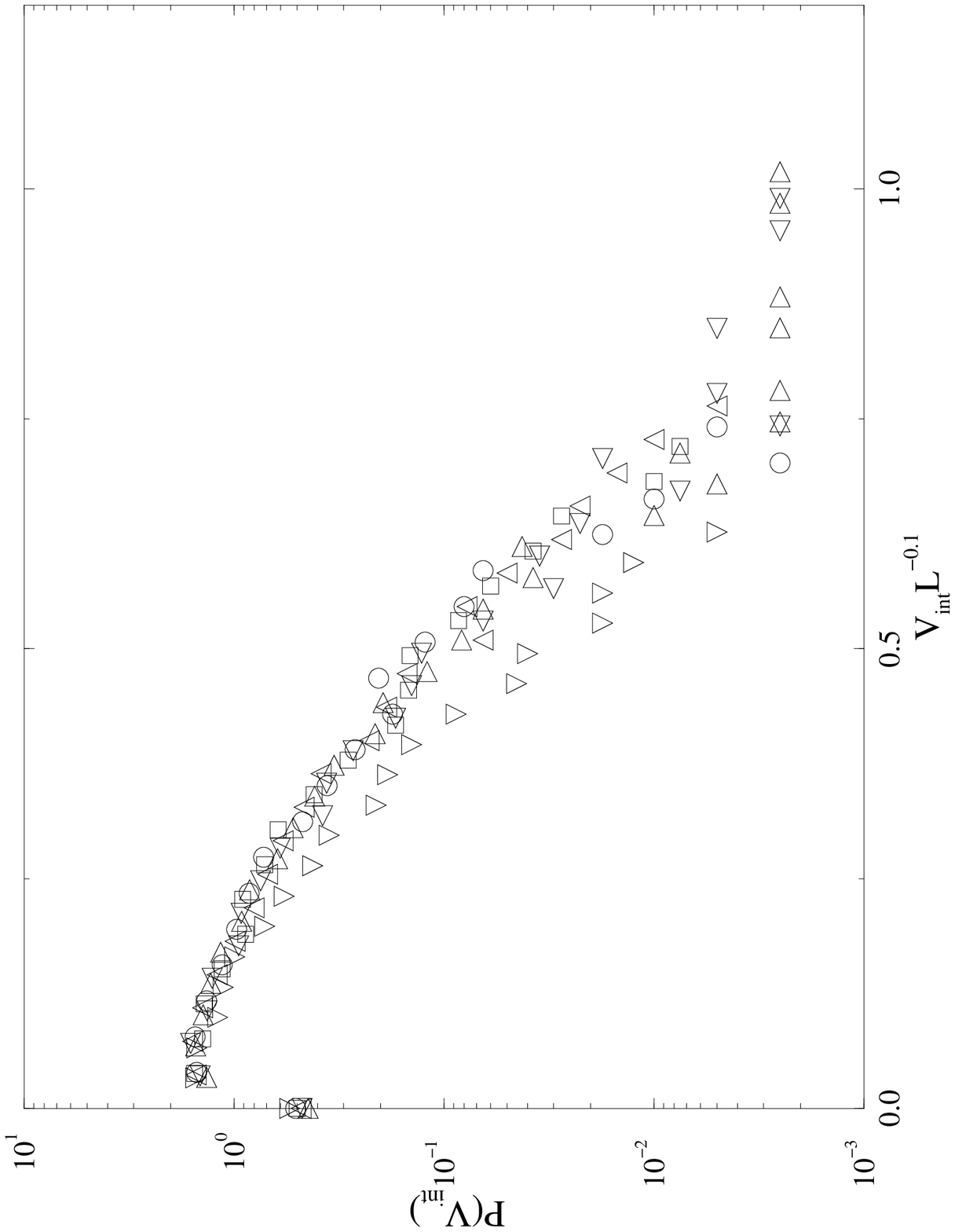}}}}
\vspace*{5mm}
\caption{
  Scaling plots of the probability distributions of $\delta E_2$ and
  $V_{int,eff}$ in 3d for the uniform distribution of the bond energies
  $e_{ij}$ in a log-log plot. 
}
\label{collapse2}
\end{figure}


\begin{references}

\bibitem{halpin} 
  T. Halpin-Healy and Y.-C. Zhang,
  Phys. Rep. {\bf 254}, 215 (1995).

\bibitem{kardar}   
  M. Kardar, G. Parisi, and Y.-C. Zhang, 
  Phys. Rev. Lett. {\bf 56}, 889 (1986).

\bibitem{mezard} 
  M. Mezard, J. Physique {\bf 51}, 1831 (1990).

\bibitem{natter} 
  T. Nattermann, M. Feigel'man, I Lyuksyutov,
  Z. Phys. B {\bf 84}, 353 (1991).

\bibitem{tang} 
  L. H. Tang, 
  J. Stat. Phys. {\bf 77}, 581 (1994).

\bibitem{muk} 
  S. Mukherji, 
  Phys. Rev. E {\bf 50}, R2407 (1994).

\bibitem{kinzel} 
  H. Kinzelbach and M. L\"assig, 
  Phys. Rev. Lett. {\bf 75}, 2208 (1995).

\bibitem{hwa} 
  T. Hwa and D. Fisher,
  Phys. Rev. B {\bf 49}, 3136 (1994).

\bibitem{oneline_num}
  M. Kardar, Phys.\ Rev.\ Lett.\ {\bf 55}, 2235 (1985);
  D. Huse and C. L. Henley, Phys.\ Rev.\ Lett.\ {\bf 54}, 2708 (1985);
  M. Kardar, Phys.\ Rev.\ Lett.\ {\bf 55}, 2923 (1985).

\bibitem{oneline}
  M.\ Kardar and Y.-C.\ Zhang, Phys.\ Rev.\ Lett.\ {\bf 58}, 2087 (1987);
  T.\ Nattermann and R.\ Lipowski, Phys.\ Rev.\ Lett. {\bf 61}, 2508 (1988);
  J.\ Derrida and H.\ Spohn, J.\ Stat.\ Phys.\ {\bf 51}, 817 (1988);
  G.\ Parisi, J.\ Physique {\bf 51}, 1695 (1990);
  D.\ Fisher and D.\ Huse, Phys.\ Rev.\ B {\bf 43}, 10728 (1991).


\bibitem{heiko}
  H.\ Rieger,  Phys. Rev. Lett. {\bf 81}, 4488 (1998).

\bibitem{flows}
  H.~Rieger, {\em Frustrated Systems: Ground State Properties via
  Combinatorial Optimization}, Lecture Notes in Physics 501 
  (Springer Verlag Heidelberg, 1998);
  R.~Ahuja, T.~Magnanti and J.~Orlin, 
  {\em Network Flows}, (Prentice Hall, New Jersey, 1993).
 
\bibitem{review} For a review on applications of these and related
  techniques to physical problems see:
  M.\ Alava, P.\ Duxbury, C.\ Moukarzel and H.\ Rieger:
  {\it Combinatorial optimization and disordered systems},
  ``Phase Transition and Critical Phenomena''
  (ed.\ C. Domb and J. L. Lebowitz), Academic Press, (2000).


\end{references}
\end{document}